\documentclass{article}
\usepackage[utf8]{inputenc}
\usepackage[parfill]{parskip}
\usepackage[margin=1.5in]{geometry}
\usepackage{amsmath,amssymb,mathtools,amsthm}
\usepackage{comment}
\usepackage{cite}
\usepackage{subcaption}
\usepackage[nottoc]{tocbibind}
\DeclareMathOperator{\E}{\mathbb{E}}

\DeclareMathOperator*{\diag}{diag}

\DeclareMathOperator*{\cov}{cov}

\DeclareMathOperator*{\var}{var}
\DeclareMathOperator*{\tr}{tr}

\DeclareMathOperator*{\argmax}{arg\ max}

\newcommand{\bs}[1]{\boldsymbol{#1}}
\usepackage{hyperref}
\hypersetup{
    colorlinks=true,
    linkcolor=blue,
    filecolor=magenta,      
    urlcolor=cyan,
    citecolor=blue,
}
\usepackage{graphicx}
\usepackage{natbib}
\usepackage{authblk} 
\newtheorem{theorem}{Theorem}[section]

\newtheorem{lemma}[theorem]{Lemma}

\title{Local False Sign Rate and the Role of Prior Covariance Rank in Multivariate Empirical Bayes Multiple Testing}
\author{Dongyue Xie}
\affil{Department of Statistics \\The University of Chicago}

\date{}

\begin{document}

\maketitle

\begin{abstract}
This paper investigates the relationship between the rank of the prior covariance matrix and the local false sign rate (lfsr) in multivariate empirical Bayes multiple testing, specifically within the context of normal mean models. We demonstrate that using low-rank covariance matrices for the prior results in inflated false sign rates, a consequence of rank deficiency. To address this, we propose an adjustment that mitigates this inflation by employing full-rank covariance matrices. Through simulations, we validate the effectiveness of this adjustment in controlling false sign rates, thereby improving the robustness of empirical Bayes methods in high-dimensional settings. Our results show that the rank of the prior covariance matrix directly influences the accuracy of sign estimation and the performance of the lfsr, with significant implications for large-scale hypothesis testing in statistics and genomics.
\end{abstract}

\section{Introduction}

Empirical Bayes (EB) methods offer a simple framework for large-scale simultaneous hypothesis testing. In a typical setting, a large number of independent test statistics (e.g., z-scores or p-values) are observed, and one of the goals is to find a subset of hypotheses that are scientifically interesting for further study. Suppose $N$ independent z-scores are observed, each with an unknown mean $\mu_i$, and $\mu_i$ is drawn independently and identically distributed (iid) from a distribution with density $g(\cdot)$,

\begin{equation} x_i\sim N(\mu_i,1), \quad \mu_i\sim g(\cdot), \end{equation}

an empirical Bayes procedure estimates $g$ by maximizing the marginal likelihood of $x_i$ and performs posterior inference of $\mu_i$ with $\hat{g}$ plugged in. EB has been widely applied in large-scale multiple testing \citep{efron2004large, stephens2017false}, shrinkage estimation \citep{johnstone2004needles,koenker2014convex,dey2018new}, and other tasks such as sparse regression and model selection \citep{martin2017ber,kim2024flexible,zou2024fast}.

The selection of a subset of hypotheses is typically based on some posterior measures of the effect, one of which is the local false sign rate (lfsr). Introduced by \citet{stephens2017false}, the lfsr is the posterior probability that we would make an error in the sign of an effect if we were to claim it as positive or negative. An analogy to lfsr is the local false discovery rate (lfdr), introduced by \citet{efron2001empirical}. The lfdr is the posterior probability of an effect being a false discovery if we were to declare it as a discovery.

Most of the methodological and theoretical development of empirical Bayes multiple testing procedures has focused on the univariate case, where each observed test statistic is one-dimensional. \citet{urbut2019flexible} extended it to multivariate settings, allowing each observed variate to be a random vector. The authors proposed multivariate adaptive shrinkage (mash), which assumes a scale mixture of multivariate normal distributions on the effect and illustrated the approach with an eQTL study. \citet{liu2024pmash} extended mash to allow Poisson-distributed observations.

\citet{urbut2019flexible} suggested using lfsr in the mash model. We have found that mash gives severely inflated false positives when claiming the sign of conditions. Further investigation suggests that this is due to a (likely undesired) property of lfsr in multivariate normal distributions and is related to the rank of the prior covariance matrix. In the following sections, we first review lfsr and mash, illustrate the issue with lfsr, and propose a solution. Simulations are performed to show the effectiveness of the solution.

\section{Review of lfsr and mash}

In this section we briefly review local false sign rate and the mash model. 
\subsection{Local false sign rate}

The local false sign rate is a posterior measure of how confident we can be in the direction of an effect. \citet{stephens2017false} defines lfsr as 
\begin{equation}\label{def:lfsr_stephens}
    lfsr = \min\{p(\mu\leq 0|x), p(\mu\geq 0|x)\},
\end{equation}
where $x$ is the observed statistics, $x|\mu\sim f(\mu)$ and $\mu$ is the unknown effect, $\mu\sim g(\cdot)$. If we were to declare the sign of effect $\mu$, our best guess would be based on the posterior probabilities $p(\mu < 0|x), p(\mu > 0|x), p(\mu=0|x)$. 



A connection between lfsr and lfdr is that a smaller value of lfsr indicates more certainty of the effect direction hence making it less likely that the effect is 0. The lfsr improves upon lfdr by recognizing that in real applications the effect $\mu$ is almost never exactly 0. Analogous to the tail FDR, \citet{stephens2017false} defines false sign rate, $FSR(\Gamma)$ as the average sign error rates over subset $\Gamma$, with an estimate given by
\begin{equation}\label{FSR_hat}
    \widehat{FSR}(\Gamma) = (1/|\Gamma|)\sum_{j\in\Gamma}lfsr_j.
\end{equation} 
Analogous to the q-value, the s-value is defined as $s_j = \widehat{FSR}(\{k:lfsr_k\leq lfsr_j\})$. The false sign proportion (FSP) is defined as
\begin{equation}\label{FSP}
    FSP(\Gamma) = \frac{V}{|\Gamma|},
\end{equation}
where $\Gamma$ is an index set in which the signs of corresponding effects are declared, and $V$ is the number of falsely claimed signs in set $\Gamma$, $V = \#\{i\in\Gamma:\text{sign}(\mu_i)\neq \widehat{\text{sign}}(\mu_i)\}$. We would expect that when using the $\widehat{FSR}$ as an estimate of false sign proportion, the false sign rate is controlled. 

\subsection{Multivariate adaptive shrinkage}

Suppose some large number $N$ of $R$-dimensional multivariate normal distributed variables $\bs x_i$ are observed, each with its own unknown mean $\bs\mu_i$, and known covariance matrix $\bs V_i$. Mash assumes a multivariate normal mixture prior on $\bs \mu_i$, 
\begin{equation}\label{model:mash}
    \begin{split}
        &\bs x_i|\bs\mu_i \sim N(\bs \mu_i,\bs V_i),
        \\
        &\bs\mu_i\sim\sum_{k=1}^K \pi_k N(\bs 0,\bs U_k).
    \end{split}
\end{equation}


The normal prior yields exact marginal distribution of $\bs x_i$ and posterior of $\bs\mu_i$, 
\begin{equation}\label{mash:marginallik}
    \bs x_i \sim \sum_k \pi_k N(\bs 0, \bs U_k+\bs V_i),
\end{equation}
\begin{equation}
    \bs\mu_i|\bs x_i \sim \sum_k\tilde{\pi}_{ik} N(\tilde{\bs\mu}_{ik},\tilde{\bs U}_{ik}),
\end{equation}
where 
\begin{equation}
\begin{split}
   & \tilde{\pi}_{ik} = \frac{\pi_k N(\bs x_i;\bs 0, \bs U_k+\bs V_i)}{\sum_l \pi_l N(\bs x_i;\bs 0, \bs U_l+\bs V_i)},
   \\
   & \tilde{\bs U}_{ik} = (\bs U_k^{-1} + \bs V_i^{-1})^{-1} = \bs U_k(\bs U_k + \bs V_i)^{-1}\bs V_i,
   \\
   & \tilde{\bs\mu}_{ik} = (\bs U_k^{-1} + \bs V_i^{-1})^{-1}\bs V_i^{-1}\bs x_i = \bs U_k(\bs U_k + \bs V_i)^{-1}\bs x_i.
\end{split}
\end{equation}

The calculation of lfsr for each effect $\mu_{ir}$ is based on 

\begin{equation}
    p(\mu_{ir}\leq 0|\bs x_i) = \sum_k\tilde{\pi}_{ik} \Phi\left(-\tilde{\mu}_{ik,r}/\sqrt{\tilde{ U}_{ik,rr}}\right),
\end{equation}
where $\tilde{\mu}_{ik,r}$ is the $r$th element of $\tilde{\bs\mu}_{ik}$ and $\tilde{ U}_{ik,rr}$ is the $r$th diagonal entry of the posterior covariance matrix.

In this paper we assume all diagonal elements of each true $\bs U_k$ are positive so none of the effects is exactly 0. The lfsr then can be at most 0.5 when randomly guessing the sign. If the diagonal of $\bs U_k$ can be 0 then the true lfsr can be at most 1 but the estimated lfsr is still upper bounded by 0.5 since the estimated covariance matrices do not have 0 diagonal entries.   

\textit{Remark}. The mash model proposed in \citet{urbut2019flexible} has additional known scaling parameters in the prior, $\bs\mu_i\sim\sum_k\sum_l \pi_{kl} N(\bs 0,\omega_l\bs U_k)$. The unknown prior covariance matrices are first estimated according to model (\ref{model:mash}) via an EM algorithm. Then the scaling parameters are added to the prior and the mixture weights are estimated, with estimated covariance matrices being fixed and additional canonical covariance matrices. For simplicity, we will focus on the model (\ref{model:mash}) where $\bs U_k$'s are all unknown.


\section{Lfsr and the rank of prior covariance matrix}

The assumption of the rank of prior covariance matrices is not explicitly specified in the mash model. One problem we have encountered is that when the posterior weights of samples concentrate on a mixture component with low rank covariance matrix, then the FSR is not controlled at the target level, particularly when the covariance matrix is rank-1, and the lfsr of effects across conditions tends to be very similar to each other. The lfsr fails to estimate the false sign proportion and leads to inflated FSR. The example below illustrates the issue and will be frequently examined throughout the paper.

\begin{figure}
    \centering
    \includegraphics[scale=0.4]{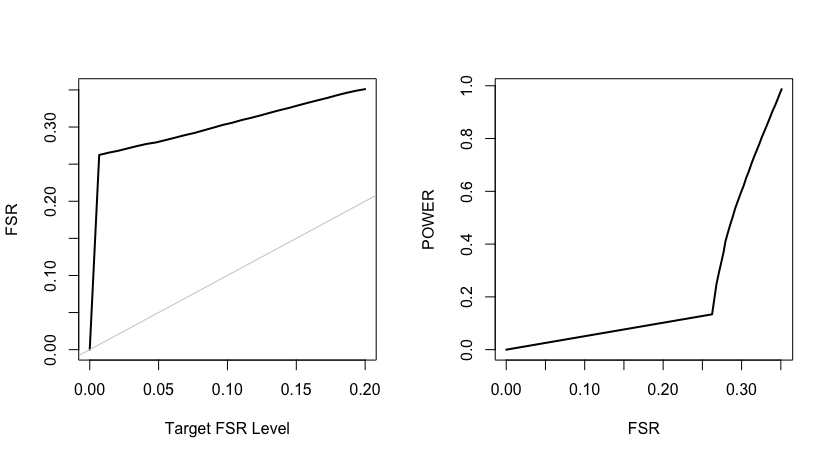}
    \includegraphics[scale = 0.4]{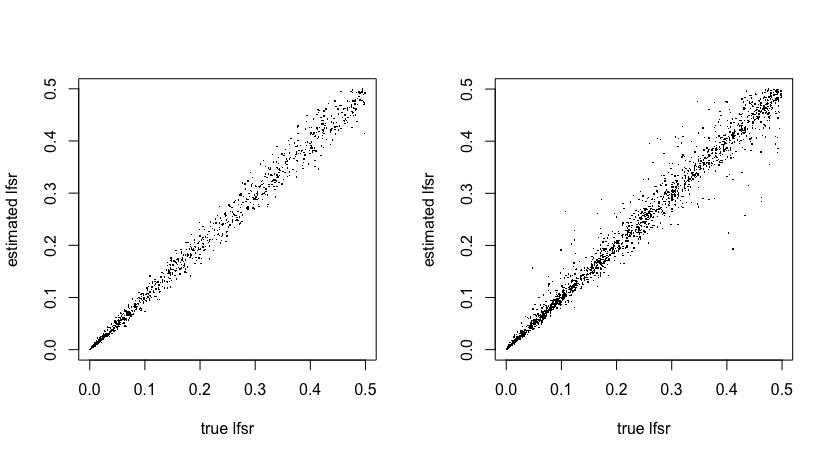}
    \caption{\textbf{Top}: Inflation of the false sign rate (FSR) when using an estimated rank-1 prior.  
\textbf{Bottom}: Comparison of true and estimated local false sign rates (lfsr) in two randomly selected runs.
}
    \label{fig:rank1problem}
\end{figure}

We generated $N=1000$ effects $\bs\mu_i$ of dimension $R=5$ from a multivariate normal distribution with covariance matrix $\bs u\bs u^T$, where $\bs u = (1,1,0.01,0.01,0.01)^T$. The samples are then generated from $\bs x_i\sim N(\bs\mu_i,\bs I)$. A mash model is fitted with two rank-1 covariance matrices estimated using ED algorithm. The point-mass and canonical covariance matrices were not included. The procedure was repeated for 30 times. The sign of effect is estimated as the sign of its posterior mean. For each repetition, we calculate false sign proportion(\ref{FSP}) and $\widehat{FSR}$(\ref{FSR_hat}).  The nominal and empirical FSR levels are compared. As shown in Figure \ref{fig:rank1problem} (top panel), the FSR is severely inflated, especially at a small target level and the size of set $\Gamma$ (see definition (\ref{FSP}), referred to as power in the plot) is very small at some desired empirical FSR level. For example, the power is less than $0.1$ at an empirical FSR level $0.1$. Figure \ref{fig:rank1problem} (bottom panel) shows the estimated lfsr and true lfsr in two randomly selected runs. The true lfsr is obtained by using true prior covariance matrices. The estimated lfsr is actually close to the true ones but the FSR is not controlled. We will later show that the estimated lfsr fails to measure the uncertainty in the estimated sign based on posterior mean.


\subsection{Rank-1 prior covariance matrix}
\label{rank1prior}

We consider a simple model where the prior is a single multivariate normal distribution with a known rank-1 covariance matrix, $\bs U = \bs u\bs u^T$, and $\bs u^T\bs u = 1$. The model for one sample is
\begin{equation}\label{mod:simple_model}
\begin{split}
     &\bs x|\bs\mu\sim N(\bs\mu,\bs I),
     \\
     & \bs \mu \sim N(\bs 0, \bs u\bs u^T).
\end{split}
\end{equation}
Under this simple model, we have the following posterior statistics:
\begin{equation}\label{rank1:post}
\begin{split}
    &\bs\mu|\bs x \sim N\left(\frac{1}{2}\bs u\bs u^T\bs x,\frac{1}{2}\bs u\bs u^T\right),
    \\
    &\mu_{r}|\bs x \sim N\left(\frac{1}{2}u_{r} (\bs u^T\bs x), \frac{1}{2}u_r^2\right),
    \\
    &p( \mu_{r}\leq 0|\bs x) = \Phi\left(-\sqrt{\frac{1}{2}}\frac{u_r(\bs u^T\bs x)}{\sqrt{u_r^2}}\right)= \Phi\left(-\sqrt{\frac{1}{2}}\frac{u_r}{|u_r|}\bs u^T\bs x\right).
\end{split}
\end{equation}

The local false sign rate of the $r$th effect $\mu_r$ is 
\begin{equation}\label{lfsr:rank1}
\begin{split}
    lfsr_{r} &= \min\left\{\Phi\left(-\sqrt{\frac{1}{2}}\frac{u_r}{|u_r|}\bs u^T\bs x\right),\Phi\left(\sqrt{\frac{1}{2}}\frac{u_r}{|u_r|}\bs u^T\bs x\right)\right\}
    \\
    &=\min\left\{\Phi\left(-\sqrt{\frac{1}{2}}\bs u^T\bs x\right),\Phi\left(\sqrt{\frac{1}{2}}\bs u^T\bs x\right)\right\}
    \\
    &= 1 - \Phi\left(\sqrt{\frac{1}{2}}|\bs u^T\bs x|\right).
\end{split}
\end{equation}

\begin{figure}
    \centering
    \includegraphics[scale=0.5]{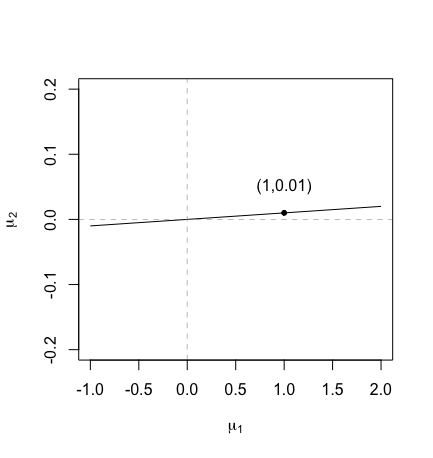}
    \includegraphics[scale=0.5]{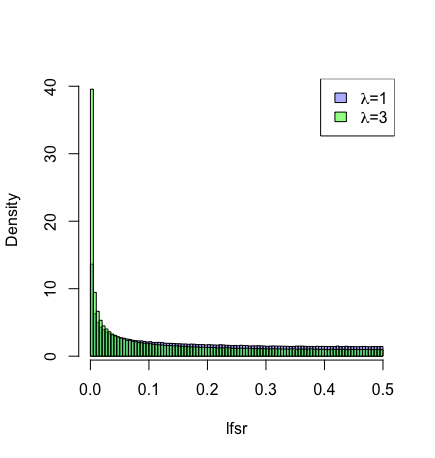}
    \caption{\textbf{Top}: Relationship between two effects in a 2D case under a rank-1 prior, illustrating the dependency structure imposed by the prior. 
\textbf{Bottom}: Distribution of the local false sign rate for different values of $\lambda$ (\(\lambda = 1\) and \(\lambda = 3\)).
}
    \label{fig:mu1mu2}
\end{figure}

The local false sign rate of $\mu_{r}$, for $r=1,2,\dots,R$, is the same regardless of $u_r$ and $x_{r}$. The rank-1 prior is very restrictive as it assumes perfect correlations among all effects. For $\bs \mu\sim N(\bs 0,\bs u\bs u^T)$, we can write $\bs\mu = \bs u z$, where $z\sim N(0,1)$. Every element of $\bs\mu$ is a multiple of a standard normal variable. As a consequence, the sign of each effect depends only on the sign of $z$. The posterior covariance matrix of $\bs \mu$ is also rank-1. Thus, even though the magnitudes of effects in $\bs\mu$ can vary considerably (due to the varying scale of $u_r$), their signs are constrained to be the same. 

The first plot in Figure \ref{fig:mu1mu2} shows the relationship between $\mu_1$ and $\mu_2$ in a two-dimensional case with a rank-1 prior. Although the effect $\mu_1$ is very small and can be arbitrarily close to the x-axis, due to the constraint of the prior, its sign is always the same as the sign of $\mu_2$. The sign of $\mu_2$ can be inferred with more certainty from the observations, so as a consequence, both effects have small lfsr.




The lfsr (\ref{lfsr:rank1}) depends only on a linear combination of $\bs x$, and the combination coefficients are given by $\bs u$. Considering marginally $\bs x\sim N(\bs 0, \bs u\bs u^T+\bs I)$, the combined variable $\bs u^T\bs x$ is distributed as $N(0,2)$. Again, this assures that when using a rank-1 prior, the lfsr is only determined by an observation from a univariate normally distributed variable. More generally, if $\bs\mu\sim N(\bs 0, \lambda\bs u\bs u^T)$ where $\lambda>0$ and $||\bs u||=1$, then $\bs u^T\bs x\sim N(0,\lambda+1)$, and

\begin{equation}
    lfsr_r =1 - \Phi\left(\sqrt{\frac{\lambda}{\lambda+1}}|\bs u^T\bs x|\right).
\end{equation}

Let $\bs u =(u_1,u_2)^T = (\lambda,a)^T/\sqrt{\lambda^2+a^2}$, where $a$ is a small constant. As $\lambda$ grows, the variance of the first effect $\var(\mu_1) = \lambda u_1^2 = \frac{\lambda^3}{\lambda^2+a^2}$ increases, while the variance of the second effect $\var(\mu_2) =\lambda u_2^2 = \frac{a^2}{\lambda+a^2/\lambda}$ approaches 0. Meanwhile, $\sqrt{\lambda/(1+\lambda)}$ approaches 1, so the lfsr is determined by the size of $\bs\mu^T\bs x$, whose variance is dominated by $\lambda$. Consequently, the effect $\mu_2$ can be arbitrarily small as $\lambda$ increases, while its lfsr is more likely to be close to 0. Figure \ref{fig:mu1mu2}, second panel, shows the distribution of lfsr when $\lambda = 1$ and $\lambda=3$. The density concentrates more near 0 when $\lambda=3$.

We have derived the lfsr under the rank-1 prior and explicitly followed the modeling assumptions. If the effects are truly generated from a multivariate normal distribution with a rank-1 covariance matrix, their lfsr are supposed to be the same. But is it a desired property in real applications? In the above example, though their lfsr are the same, it's tempting to say that the second effect $\mu_2$ is more likely to be a null, while the first one could be a more interesting discovery. 

The prior has been assumed known so far, and we now consider the prior covariance matrices being estimated from the samples. Recall in Figure \ref{fig:rank1problem}, the effects were generated from a prior with rank-1 covariance matrices, and we used the estimated rank-1 matrices for posterior calculation. The estimated and true lfsr are highly correlated, but the FSR is still inflated. Under the model (\ref{mod:simple_model}), assume the estimated prior covariance matrix is normally distributed with a covariance matrix of order $1/n$,
\begin{equation}
    \begin{split}
        &\hat{\bs u} = \bs u + \bs\epsilon,
        \\
        &\bs\epsilon\sim N\left(0,\frac{1}{n}\bs \Sigma\right).
    \end{split}
\end{equation}
Without loss of generality, we assume $\bs u^T\bs x<0$, then the true lfsr is then $\Phi(\sqrt{1/2}\bs u^T\bs x)$. The estimated lfsr is 
\begin{equation}
\begin{split}
    \widehat{lfsr} &= \Phi\left(\sqrt{\frac{1}{2}}\hat{\bs u}^T\bs x\right)
    \\
    &=\Phi\left(\sqrt{\frac{1}{2}}\bs u^T\bs x + \sqrt{\frac{1}{2}}\bs\epsilon^T\bs x\right) 
    \\
    &= \Phi\left(\sqrt{\frac{1}{2}}\bs u^T\bs x + \sqrt{\frac{\bs x^T\bs \Sigma\bs x}{2n}}z\right),
\end{split}
\end{equation}
where $z\sim N(0,1)$ and we have treated $\bs x$ as given and $\bs\epsilon$ as random. Taking the expectation of $\widehat{lfsr}$ with respect to $z$, we have
\begin{equation}
\begin{split}
    \E_z\left(\widehat{lfsr}\right) &= \E_z \Phi\left(\sqrt{\frac{\bs x^T\bs \Sigma\bs x}{2n}}z +\sqrt{\frac{1}{2}}\bs\mu^T\bs x \right)
    \\
    &= \Phi\left(\frac{\sqrt{\frac{1}{2}}\bs\mu^T\bs x}{\sqrt{1+\frac{\bs x^T\bs \Sigma\bs x}{2n}}}\right).
\end{split}
\end{equation}

Since $\bs x^T\bs \Sigma\bs x\geq 0$, the denominator $\sqrt{1+\frac{\bs x^T\bs \Sigma\bs x}{2n}}$ is at least 1, and we conclude $\E\left(\widehat{lfsr}\right)\geq lfsr$. The estimated lfsr is expected to overestimate the true lfsr, particularly when elements of $\bs x$ are large, so $\sqrt{1+\frac{\bs x^T\bs \Sigma\bs x}{2n}}$ is much larger than 1. In practice, this is not a big issue as long as $\hat{\bs u}^T\bs x$ and $\bs u^T\bs x$ are of the same sign, and we observe that the estimated and true lfsr give very similar rejection set $\Gamma$ at a target FSR level $\alpha$. 

We now show that, when using a rank-1 prior covariance matrix, the sign of the posterior mean is not robust to estimation error. The inflated FSR in Figure \ref{fig:rank1problem} is due to the wrong estimate of signs based on the posterior mean, while lfsr does not account for such uncertainty. Recall that the true posterior mean of $\mu_r$ is $\frac{1}{2}u_r(\bs u^T\bs x)$, and the estimated posterior mean is $\frac{1}{2}\hat u_r(\hat{\bs u}^T\bs x) = \frac{1}{2}(u_r+\epsilon_r)({\bs u}+\bs\epsilon)^T\bs x$, where $\epsilon_r\sim N(0,\sigma^2_r/n)$. 

Given $\bs x$, the signs of $\bs u^T\bs x$ and $\hat{\bs u}^T\bs x$ are the same with high probability, since 
\begin{equation}
\begin{split}
    p\left((\bs u^T\bs x)(\hat{\bs u}^T\bs x)>0\right)&=p\left((\bs u^T\bs x)\bs\epsilon^T\bs x>-(\bs u^T\bs x)^2\right)
    \\
    &=1-\Phi\left(-\sqrt{\frac{n(\bs u^T\bs x)^2}{\bs x^T\bs \Sigma\bs x}}\right).
\end{split}
\end{equation}

As $n$ gets larger, the probability gets closer to 1. But the sign of $u_r$ and $u_r+\epsilon_r$ differ with high probability when $u_r$ is small, since
\begin{equation}
    \begin{split}
        p\left(u_r(u_r+\epsilon_r)<0\right) &= p(u_r\epsilon_r<-u_r^2)
        \\
        &=\Phi\left(-\sqrt{\frac{nu_r^2}{\sigma^2_r}}\right).
    \end{split}
\end{equation}

\begin{figure}
    \centering
    \includegraphics[scale=0.4]{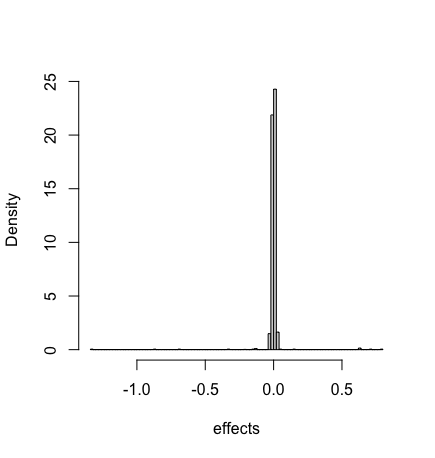}
    \includegraphics[scale=0.4]{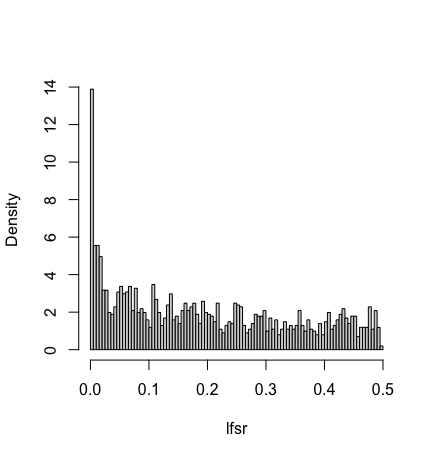}
    \caption{Histogram of effects where the estimated signs are opposite to those of the true posterior mean, along with their corresponding estimated local false sign rates.}
    \label{fig:lfsr_dist_wrongsign}
\end{figure}

When $u_r$ is small, for example, $u_r^2\ll 1/n$, the probability can be very close to $0.5$. Thus, for small effects, the estimates of their signs based on the true and estimated posterior means would disagree approximately half of the time. However, we have shown that the rank-1 prior leads to the same lfsr for all conditions, and the lfsr could be small as long as there is one large effect in any of the conditions. This implies that the lfsr of small effects is small due to the rank-1 prior constraint, while we are likely to make mistakes in claiming their signs because they are close to 0 and subject to estimation error. The FSR is then inflated, especially when the target level $\alpha$ is small. Figure \ref{fig:lfsr_dist_wrongsign} shows the distribution of effects whose signs are estimated oppositely to those of the true posterior mean and their estimated lfsr. As expected, most of the discrepancies occur when effects are small, while the lfsr does not properly reflect the uncertainty in signs, as they tend to concentrate near 0.

\subsection{Dimension of lfsr}

\begin{figure}
    \centering
    \includegraphics[scale=0.4]{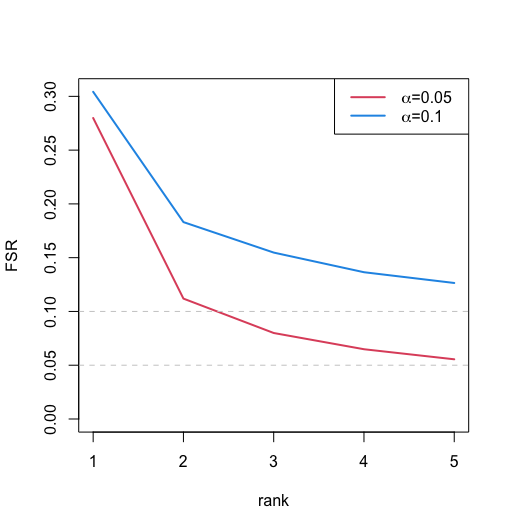}
    \caption{Effect of prior covariance matrix rank on the false sign rate (FSR) at target levels of 0.05 and 0.1. The covariance matrices are estimated using the ED algorithm, initialized with randomly generated full-rank positive definite matrices. The rank of the estimated prior covariance matrices is modified by truncating eigenvalues at zero, and the resulting rank-adjusted matrices are then used as input to the mash function.}
    \label{fig:lfsr_rank}
\end{figure}

We have shown that if the posterior weights of an observation concentrate on a normal distribution with an estimated rank-1 prior, the lfsr is likely to be invalid, especially for small effects, and consequently, the FSR is not controlled at the nominal level. A natural question to ask is what happens if the prior covariance matrix is not rank-1 but still rank-deficient. Figure \ref{fig:lfsr_rank} shows how FSR changes with the rank of the prior covariance matrix at target FSR levels of $0.05$ and $0.1$. As the rank increases, the FSR gradually drops to the nominal level. This suggests that it is preferable to use full-rank covariance matrices in the prior, even when the true prior covariance matrix is rank-1.

We define the dimension of lfsr as the number of independent dimensions it can be decomposed into and will show that its dimension equals to the rank of prior covariance matrix. Denote the eigen-decomposition of $\bs U$ as $\bs U = \sum_{k=1}^K\lambda_k \bs q_k\bs q_k^T$, where $K$ is the rank of $\bs U$, $\bs q_k^T\bs q_k = 1$, and $\bs q_k^T\bs q_{k'} = 0$ for $k\neq k'$, the posterior negative probability is
\begin{equation}\label{lfsr:rankK}
\begin{split}
    p(\mu_{r}\leq 0|\bs x) &= \Phi\left(-\frac{\sum_k\tilde{\lambda}_k q_{kr}(\bs q_k^T\bs x)}{\sqrt{\sum_k \tilde{\lambda}_k q_{kr}^2}}\right) 
    \\
    &= \Phi\left(-\sum_k \tilde{\tilde{\lambda}}_{kr} (\bs q_k^T\bs x)\right),
\end{split}
\end{equation}
where $\tilde{\lambda}_k = \frac{\lambda_k}{1+\lambda_k}$, and $\tilde{\tilde{\lambda}}_{kr} = \frac{\tilde{\lambda}_k q_{kr}}{\sqrt{\sum_k \tilde{\lambda}_k q_{kr}^2}}$.

The lfsr, as shown in (\ref{lfsr:rankK}), can be decomposed into a summation of \(K\) components. Since \(\text{cov}(\bs q_k^T\bs x, \bs q_{k'}^T\bs x) = 0\) for \(k,k' \in [K]\), each component of lfsr is independent in the sense that each one is determined by an observation from a univariate normally distributed variable that is independent of the others. The dimension of lfsr equals \(K\), the rank of \(\bs U\). If the true lfsr has dimension \(K\) while we specify a prior covariance matrix with rank smaller than \(K\), then the assumed prior imposes extra constraints on the effects, and the lfsr obtained from the fitted model would be invalid, as we have shown in Section (\ref{rank1prior}). Even if the rank of the chosen prior covariance matrix is the same as the true one but smaller than the dimension \(R\), the FSR would still be inflated, as shown in Figure (\ref{fig:lfsr_rank}). 

As mentioned above, in real applications, we might use lfsr for ordering the effects and expect the lfsr to reflect their significance. In other words, the desired dimension of lfsr, hence the rank of \(\bs U\), should equal the number of conditions \(R\), such that effects are not constrained to lower dimensions and their direction can vary freely.

\section{Full-rank prior covariance matrix}

We propose to use full-rank covariance matrices in the prior, $rank(\bs U) = R$. Note that this proposal is advised even if the true $\bs U$ is rank-deficient. The benefits of using a full-rank covariance matrix are twofold: it prevents unwanted constraints on the effects and improves robustness to estimation uncertainty. 

To ensure the covariance matrix is full rank and positive definite, we assume the following prior of $\bs \mu$ 
\begin{equation}
    \bs\mu\sim N(\bs 0, \bs U + \bs D),
\end{equation}
where $\bs U$ is a positive semi-definite matrix and $\bs D = \text{diag}\{\sigma^2_r\}_{r=1,2,...,R}$ is a diagonal matrix with $\sigma^2_r> 0$. When $\bs D = \sigma^2 \bs I$, this is equivalent to adding $\sigma^2$ to the eigenvalues of $\bs U$. Another possibility is to incorporate $\bs D$ into the eigenvalues of $\bs U$, yielding the prior covariance matrix $\bs Q(\bs \Lambda + \bs D)\bs Q^T$. For simplicity and generality, we add $\bs D$ to $\bs U$ directly.  

\textit{Example}. In this example we consider again the simple model. Let $\bs x\sim N(\bs \mu, \bs I)$,  $\bs\mu\sim N(\bs 0, \bs u\bs u^T + \sigma^2 \bs I)$ where $\bs u^T\bs u =1$. Then 
\begin{equation}
    \bs\mu|\bs x \sim N\left(\frac{(2\sigma^2\bs I + \bs u\bs u^T)\bs x}{2(\sigma^2+1)},\frac{2\sigma^2\bs I + \bs u\bs u^T}{2(\sigma^2+1)}\right),
\end{equation}

and

\begin{equation}\label{negp:fullrank}
    \begin{split}
        p(\mu_{r}\leq 0|\bs x) = \Phi\left(-\sqrt{\frac{1}{2(\sigma^2+1)}}\frac{2\sigma^2x_{r}+u_r(\bs u^T\bs x)}{\sqrt{2\sigma^2 + u_r^2}}\right).
    \end{split}
\end{equation}

\begin{figure}
    \centering
    \includegraphics[scale=0.5]{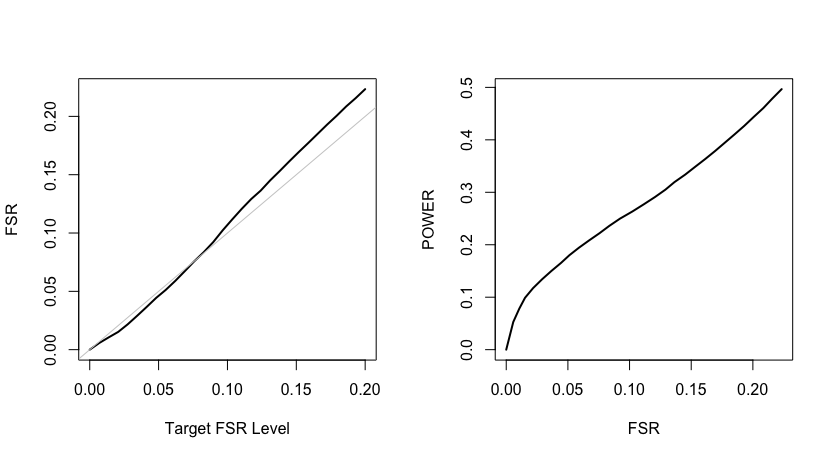}
    \caption{FSR is controlled by adding $0.03\times \bs I$ to the estimated rank-1 prior covariance matrices.}
    \label{fig:lfsr_const_adj}
\end{figure}

If $\sigma^2 = 0$, (\ref{negp:fullrank}) recovers the posterior negative probability in (\ref{rank1:post}). With $\bs D$ added, the lfsr of the $r$th condition depends on both $x_r$ and $u_r$, and the elements of $\bs\mu$ would have different lfsr given $\bs x$. Recall that the inflated FSR in the rank-1 prior case is due to the incorrect estimation of signs (of small effects) based on the estimated posterior mean, while the estimated lfsr (possibly very small) is the same for all conditions. By adding $\sigma^2\bs I$, although there are still some disagreements in signs between the estimated and true posterior means, the lfsr of small effects is now allowed to be large and hence properly reflects the uncertainty in their signs. In Figure \ref{fig:lfsr_const_adj}, the FSR is controlled by simply adding $0.03\times \bs I$ ($0.03$ is carefully chosen) to the estimated rank-1 priors, and the power increases substantially compared to the results in Figure \ref{fig:rank1problem}.

\subsection{Choose $\bs D$}

The adjustment $\bs D$ is preferably data-adaptive instead of being fixed. An immediate way to determine $\bs D$ is to estimate it from data along with $\bs U$. However, if $\bs U$ is full-rank, then $\bs U$ and $\bs D$ are not identifiable because we can always split $\bs D$ into two additive matrices and add one matrix to $\bs U$, while the distribution of $\bs x$ remains the same. If $\bs U$ and $\bs D$ are to be estimated, we need to assume that the rank of $\bs U$ is smaller than its dimension $R$ for the identification of both matrices.

The diagonal matrix $\bs D$ together with $\bs U$ can be estimated using maximum likelihood estimators. The estimated covariance matrix $\widehat{\bs U}^{mle}$ and $\widehat{\bs D}^{mle}$ maximize the log-likelihood 
\begin{equation}
    l(\bs U,\bs D;\bs x) = -\frac{1}{2}\sum_i\log|\bs V_i+\bs U + \bs D|-\frac{1}{2}\sum_i\bs x_i^T(\bs V_i+\bs U + \bs D)^{-1}\bs x_i.
\end{equation}

However, in practice, the diagonal elements of $\bs D$ can be estimated very close to or exactly equal to 0. Then, the covariance matrix $\widehat{\bs U}^{mle} + \widehat{\bs D}^{mle}$ is (almost) rank-deficient. 

\textit{Example}. Assume $\bs V_i = \bs I$ for all samples, $\bs U = \bs u\bs u ^T$ where $||\bs u|| = 1$, and $\bs D = \sigma^2\bs I$. Then, $\bs x_i\sim N(\bs 0, \bs u\bs u^T+(1+\sigma^2)\bs I)$, and the MLE of $\sigma^2$ is 
\begin{equation}
    \hat\sigma^2 = \left(\frac{\sum_{r=2}^R \lambda_r}{R-1} - 1\right)_+,
\end{equation}
where $\lambda_r$ is the $r$th largest eigenvalue of the sample covariance matrix and $(a)_+ = \max(a,0)$. As long as the mean of $\lambda_r$ for $r=2,...,R$ is smaller than 1, the estimated $\sigma^2$ would be 0. This is very likely to happen when $\bs x_i$ is generated from a degenerate multivariate normal distribution or when the sample size is small.

To resolve this issue, we can estimate $\bs D$ as large as possible while remaining consistent with the data. In other words, we choose $\widehat{\bs D}$ to be the upper bound of its interval estimation, and the estimator is denoted as $\widehat{\bs D}^{ub}$. Then, we proceed to plug in $\widehat{\bs U}^{mle} + \widehat{\bs D}^{ub}$ in the model and calculate posterior quantities.

To estimate $\bs D$ using MLE, we have to assume $\bs U$ is low-rank. Additionally, to obtain uncertainty quantification, $\sigma^2_r, r=1,2,...,R$ has to be interior points of the parameter space, i.e., $\sigma^2_r>0$, but this might not always be the case. To address these concerns, we can simply fit the model without $\bs D$ and estimate the diagonal of $\bs U$ to be the upper bound of its interval estimation. The diagonal of matrix $\bs D$ can be regarded as the difference between $\text{diag}(\widehat{\bs U})$ and its upper bound of the confidence interval. Since the diagonal elements of $\bs U$ are always positive, they are interior points of the parameter space, and the interval estimation would be valid.

Based on the asymptotic normality of MLE under regular conditions, we can evaluate the inverse Fisher information to obtain the standard errors of $\widehat{\bs U}^{mle}$.

Let $\bs T_i = \bs U + \bs V_i$, and $\bs\theta = (\text{vec}(\bs U))^T$, where $\text{vec}(\bs U)$ is the vectorized $\bs U$ and contains only unique elements, so $\text{vec}(\bs U)\in\mathbb{R}^{R(R+1)/2}$. Define $l(\bs\theta) = \log L(\bs\theta)$ as the log-likelihood. Note that elements of $\bs T_i$ are functions of $\bs\theta$, and for simplicity, we write $\bs T_i$ instead of $\bs T_i(\bs\theta)$. We denote the (expected) Fisher information matrix as $\bs I(\bs\theta)$, a symmetric matrix with $(R(R+1)/2)$ columns, whose $(j,k)$th entry is 
\begin{equation}
    \bs I_{jk}(\bs\theta) = \frac{1}{2}\sum_i\tr\left(\bs T_i^{-1}\frac{\partial\bs T_i}{\partial\theta_j}\bs T_i^{-1}\frac{\partial\bs T_i}{\partial\theta_k}\right).
\end{equation}

The asymptotic variance of $\hat{\bs\theta}^{mle}$ is given by the inverse Fisher information $\bs I(\bs\theta)^{-1}$, which can be estimated by $\bs I(\hat{\bs\theta})^{-1}$. Detailed derivations of $\bs I(\bs\theta)$ are given in Appendix \ref{appendix:infomat}. A frequently used alternative to the expected Fisher information is the observed information $\bs J(\bs\theta)$. The observed information is simply the negative second-order derivative of the log-likelihood function and is usually easier to derive. It turns out that for the multivariate normal distribution, the expected Fisher information is easier to derive and compute than the observed information.

Based on the asymptotic confidence interval, we choose our estimate of $u_{rr}$, the $r$th diagonal element of $\bs U$, to be 
\begin{equation}
    \hat{u}_{rr}^{ub} = \hat{u}_{rr}^{mle} + z_{1-\alpha/2}s.e.(\hat{u}_{rr}^{mle}),
\end{equation}
where $s.e.(\hat{u}_{rr}^{mle})$ is the standard error, given by the square root of the corresponding entry of $\bs I(\hat{\bs\theta})^{-1}$.

For additional simplicity, we develop a lower bound for the variance of $\text{diag}(\widehat{\bs U})$ and set the upper bound of the confidence interval as a multiple of the square root of the variance lower bound. The lower bound represents the minimum variance of the estimator when $\text{diag}(\widehat{\bs U}) = \bs 0$. It avoids calculating the Fisher information matrix and its inverse, making it faster when $R$ is large. See Appendix \ref{appendix:lb} for further details.


\begin{figure}
    \centering
    \includegraphics[scale=0.5]{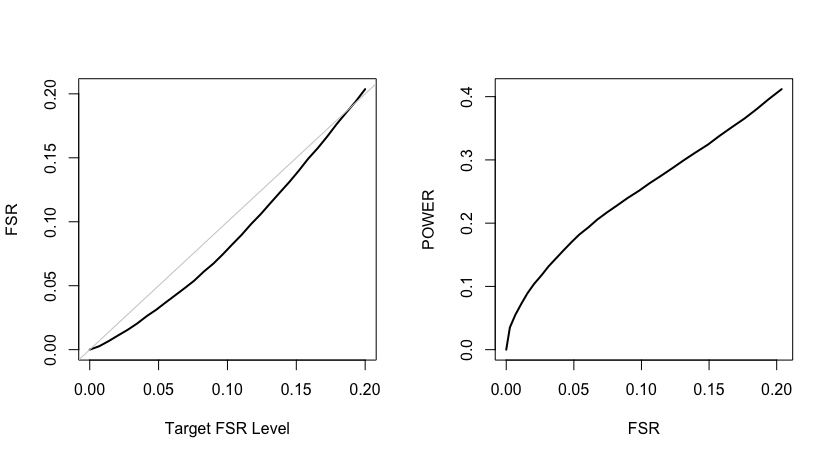}
    \includegraphics[scale=0.5]{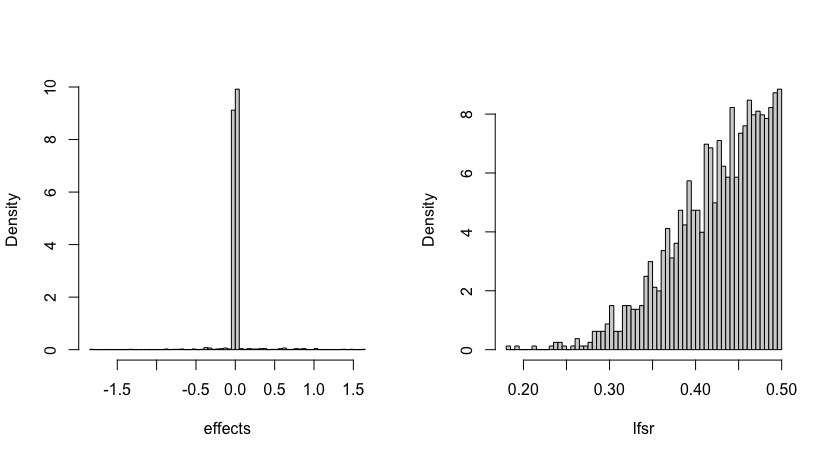}
    \caption{\textbf{Top}: the covariance  matrices are adjusted to be full-rank using the information matrix approach. \textbf{Bottom}: distribution of true effects whose signs are estimated oppositely by true and estimated posterior mean, and distribution of their estimated lfsr}
    \label{fig:lfsr_infomat}
\end{figure}

\section{Mixture model}

Our discussion has been focusing on a single normal prior, and in this section, we consider the model with a mixture prior. Each mixture component should have a full-rank covariance matrix; otherwise, the lfsr of each condition is again constrained, and the FSR cannot be controlled. We assume the mixture prior has the following form,

\begin{equation}
    \bs \mu_i\sim \sum_{k=1}^K\pi_k N(\bs 0, \bs U_k+\bs D_k),
\end{equation}

where each $\bs U_k$ is positive semidefinite, and $\bs D_k$ is a positive definite diagonal matrix. The estimated (profile) likelihood confidence interval via the \textit{rule of 2} still applies in such cases, and we can set $\text{diag}(\widehat{\bs U}_k)$ to be the upper bound of their confidence intervals. One concern is that there are a total of $KR(R+1)/2+K$ parameters in the mixture model. The profile likelihood is much more computationally intensive to evaluate. The estimated likelihood confidence interval of a single parameter ignores the uncertainty of all other parameters. This might give a very narrow confidence interval and would not address the issue of lfsr discussed above.

The calculation of the Fisher information matrix is nontrivial because of the difficulty in differentiating the marginal log-likelihood,

\begin{equation}
    l(\bs \pi,\bs U;\bs x) = \sum_i\log\left(\sum_k \pi_k N\left(\bs x_i;\bs 0, \bs V_i+\bs U_k\right)\right),
\end{equation}

as well as the large number of parameters in the mixture model.

The marginal log-likelihood is usually maximized by the EM algorithm, and we consider calculating the information matrix within the EM framework. The goal is to find the Fisher information of parameters $\bs \theta = \{\bs \pi_k,\bs U_k\}_{k=1}^K$. The observed data $\bs x$ is usually referred to as incomplete data, and the unobserved mixture component assignment (usually called missing data) is denoted as $\bs z$. The complete data log-likelihood is $l(\bs\theta;\bs x,\bs z)$, and the incomplete data log-likelihood is $l(\bs\theta;\bs x)$. 

The EM algorithm iterates between the following two steps until a convergence criterion is met:
\begin{itemize}
    \item E-step. Evaluate the expectation of the complete data log-likelihood $\E_{\bs z|\bs x,\bs\theta^{(t)}}l(\bs\theta;\bs x,\bs z)$ at the current estimates $\bs\theta^{(t)}$. 
    \item M-step. Find $\bs\theta^{(t+1)} = \argmax_{\bs\theta} \E_{\bs z|\bs x,\bs\theta^{(t)}}l(\bs\theta;\bs x,\bs z)$.
\end{itemize}

The expected complete data log-likelihood is 
\begin{equation}\label{EM:eloglik}
    \E_{\bs z|\bs x,\bs\theta^{(t)}}l(\bs\theta;\bs x,\bs z) = \sum_{i,k} \gamma_{ik} \left(\log\pi_k + \log N(\bs x_i;\bs 0,\bs U_k+\bs V_i)\right),
\end{equation}
where $\gamma_{ik} = p(z_{ik} = 1|\bs x_i,\bs\theta^{(t)})$ is the posterior probability of mixture assignment.

There is extensive literature on finding the information matrix within the EM framework. \citet{louis1982finding} proposed the following formula to calculate the observed information,  
\begin{equation}\label{EM:louis}
\begin{split}
    \bs J(\bs\theta) = &-\E\frac{\partial^2 l(\bs\theta;\bs x,\bs z)}{\partial\bs\theta\partial\bs\theta^T}
    - \E \left(s(\bs\theta;\bs x,\bs z)s^T(\bs\theta;\bs x,\bs z)\right) + \E s(\bs\theta;\bs x,\bs z)\E s^T(\bs\theta;\bs x,\bs z),
\end{split}
\end{equation}
where $s(\bs\theta;\bs x,\bs z)$ is the score of $\bs\theta$ with complete data, and the expectation is taken with respect to $\bs z|\bs x,\bs\theta$. See \citet{oakes1999direct} for an alternative formula for calculating $\bs J(\theta)$. All of the conditional expectations in (\ref{EM:louis}) can be calculated in the final iteration of EM when it converges. However, $\bs J(\theta)$ can be a high-dimensional matrix, and calculating its inverse can pose singularity issues. It also seems redundant to perform these calculations solely to determine the variance of the diagonal elements of each $\bs U_k$.

It's tempting to simply use the expected complete data log-likelihood (\ref{EM:eloglik}) for the calculation of the Fisher information matrix. The main reason is that the parameters $\pi_k$, $\bs U_k$ for $k=1,2,\dots,K$ are all orthogonal under the expected complete data log-likelihood, so the corresponding Fisher information matrix is a block diagonal matrix. For our purpose, we can simply formulate the Fisher information for each $\bs U_k$ separately, then take its inverse to obtain the variances. We provide a justification of the approach as follows. The complete data log-likelihood can be decomposed into two parts,  
\begin{equation}\label{EM:loglik-decompose}
    l(\bs\theta;\bs x,\bs z) = \log p(\bs z|\bs x,\bs\theta) + l(\bs\theta;\bs x).
\end{equation}

Taking the expectation of both sides with respect to $\bs z|\bs x,\bs\theta$, then taking the second derivative with respect to $\bs\theta$, we obtain  
\begin{equation}
    \frac{\partial\E l(\bs\theta;\bs x,\bs z)}{\partial\bs\theta\partial\bs\theta^T} = \frac{\partial \E\log p(\bs z|\bs x,\bs\theta)}{\partial\bs\theta\partial\bs\theta^T} + H(\bs\theta;\bs x),
\end{equation}
where $H(\bs\theta;\bs x)$ is the Hessian matrix of $l(\bs\theta;\bs x)$.

Finally, taking the expectation with respect to $\bs x$ on both sides, we have  
\begin{equation}
\begin{split}
    \bs I(\bs\theta;\bs x) &= - \E \left(\frac{\partial \E_{\bs z|\bs x,{\bs\theta}}l(\bs\theta;\bs x,\bs z)}{\partial \bs\theta\partial \bs\theta^T}\right) - (-\E\frac{\partial \E_{\bs z|\bs x,\bs\theta}\log p(\bs z|\bs x,\bs\theta)}{\partial\bs\theta\partial\bs\theta^T} )
    \\
    &:= \bs I(\bs\theta;\bs x,\bs z) - \bs I(\bs\theta;\bs z|\bs x).
\end{split}
\end{equation}

Given $\bs I(\bs\theta;\bs z|\bs x)$ is positive definite, $\bs I(\bs\theta;\bs x,\bs z)>\bs I(\bs\theta;\bs x)$ and $\bs I^{-1}(\bs\theta;\bs x,\bs z)<\bs I^{-1}(\bs\theta;\bs x)$, where $\bs A<\bs B$ denotes that $\bs B-\bs A$ is positive definite. Since the information matrix is symmetric, we conclude that the diagonal elements of the inverse Fisher information based on the expected complete data log-likelihood are smaller than those obtained using the incomplete data likelihood. This is expected since we regard $\gamma_{ik}$ in (\ref{EM:eloglik}) as given, and complete data carry more information than the observed data. Thus, the uncertainty in parameter estimation decreases. The upper bound of the confidence interval based on the complete data would be smaller than the actual one. This is suitable for our case since we would like to estimate the diagonal of $\bs U_k$ to be larger than the MLE while remaining consistent with the data.

To conclude, we use the following estimated expected Fisher information matrix for constructing the confidence interval of $\text{diag}(\bs U_k)$, 
\begin{equation}
    \widehat{\bs I(\bs\theta)} = - \E \left(\frac{\partial \E_{\bs z|\bs x,\hat{\bs\theta}}l(\bs\theta;\bs x,\bs z)}{\partial \bs\theta\partial \bs\theta^T}\right)\bigg\rvert_{\bs\theta = \hat{\bs\theta}}.
\end{equation}
We then set the diagonal of each $\widehat{\bs U}_k$ to be the upper bound of their confidence intervals while the off-diagonal elements remain the same as MLE. We will show that the methods work well in practice.

Figure \ref{fig:lfsr_infomat} shows the results after adjusting the estimated covariance matrices to be full-rank using the Wald confidence interval approach. In the bottom panel, we can see that most of the sign discrepancies still occur when effects are small, and the their lfsr are now larger compared to the ones in Figure \ref{fig:lfsr_dist_wrongsign} before adjustment.

\section{Simulations}

In the first simulation, we generate effects from a mixture of two normal distributions, and the covariance of observations is an identity matrix; then, we consider a more general setting where the covariance of observations, $\bs V_i$, can vary across samples.

The two covariance matrices in the prior are $\bs u_1\bs u_1^T$, where $\bs u_1 = (1,1,0.01,0.01,0.01)^T$, and $\bs u_2\bs u_2^T$, where $\bs u_2 = (1,-1,1,0.01,0.01)^T$. Each covariance matrix is then multiplied by 3. The prior weights are the same, $\pi_k = 1/2$ for $k=1,2$. In each repetition, $N=1000$ effects are generated, and a mash model is fitted. Prior covariance matrices are specified as rank-1 and estimated by the ED algorithm. For the more general setting, $\bs V_i$ is drawn from a Wishart distribution with $df = 10$ and scale matrix $\bs I$, then the draws are divided by $10$ so that the mean is at $\bs I$ and the generated matrices are well-conditioned.

\begin{figure}
    \centering
    \includegraphics[scale=0.3]{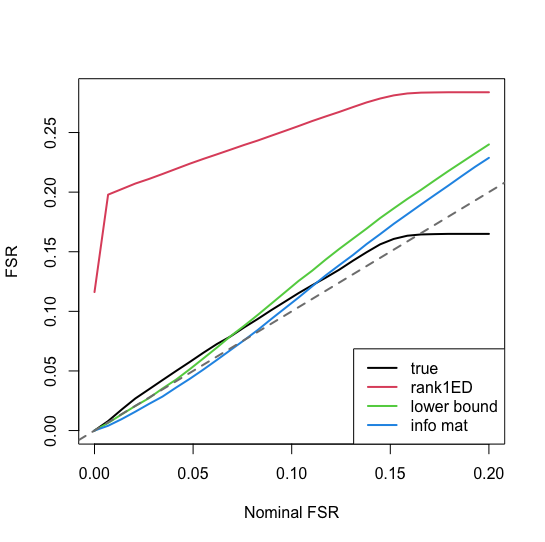}
    \includegraphics[scale=0.3]{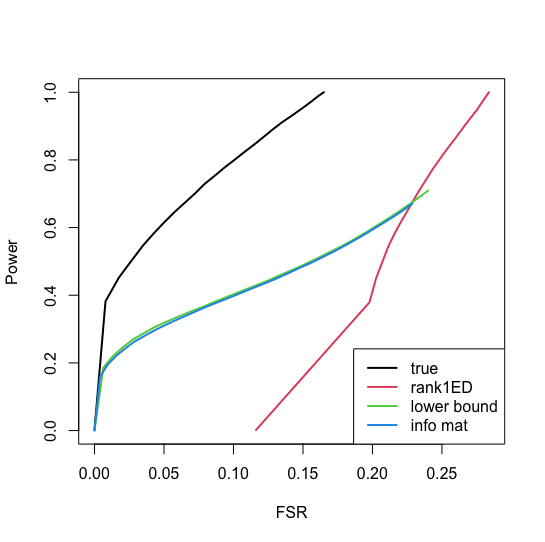}
    \includegraphics[scale=0.3]{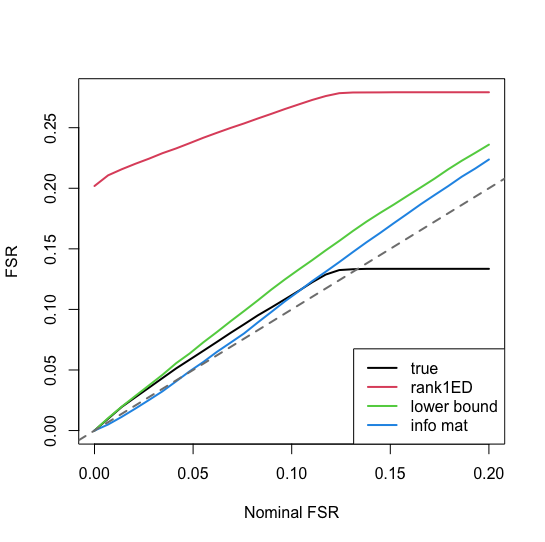}
    \includegraphics[scale=0.3]{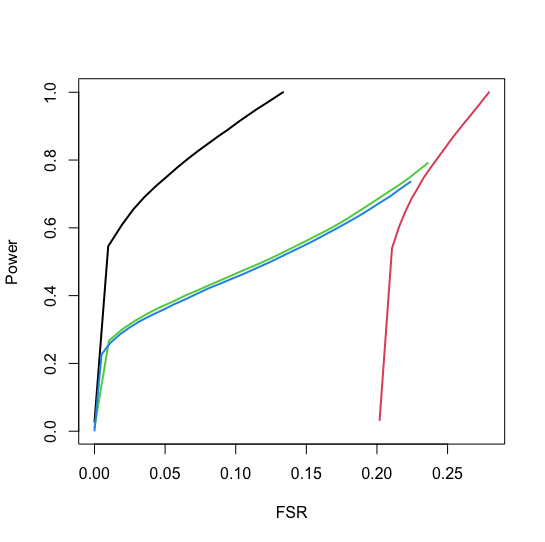}
    \caption{Simulation results comparing nominal False Sign Rate (FSR) and empirical FSR, as well as empirical FSR versus power. 
The settings include: \textbf{true} (using the true covariance matrix), \textbf{rank1ED} (using an estimated rank-1 matrix from eigen decomposition), and two full-rank adjustments—\textbf{lower bound} and \textbf{info mat}. 
The top panel represents the case where $\bs V_i = \bs I$, while the bottom panel corresponds to varying $\bs V_i$ across samples.}
    \label{fig:simu12}
\end{figure}

The results in Figure \ref{fig:simu12} suggest that by adding a diagonal matrix to the estimated rank-1 matrices, the issue of inflated FSR almost disappears. Adding the diagonal matrix derived from the information matrix gives slightly better control of FSR than the lower bound method, while the power loss is small. This is likely because the former method considers the scale of matrix $\bs U$. Compared to the results when true prior covariance matrices are given, both adjustments are able to control FSR close to the true one and at the target level. However, there seems to be power loss.

\begin{figure}
    \centering
    \includegraphics[scale=0.35]{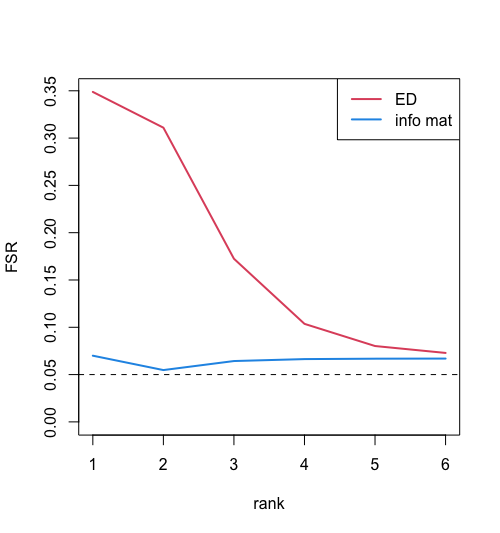}
    \includegraphics[scale=0.35]{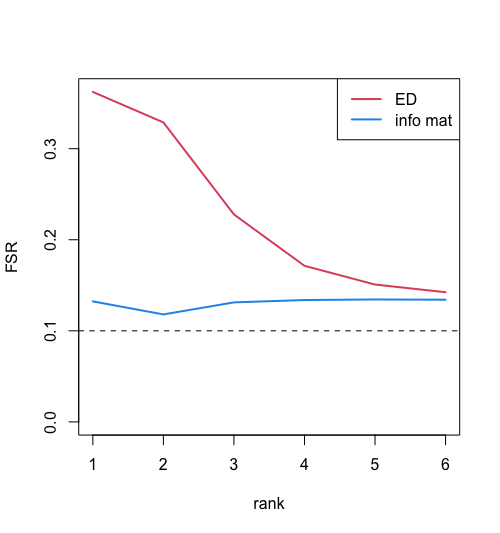}
\caption{Variation of the False Sign Rate (FSR) with the rank of estimated covariance matrices from ED. The left panel corresponds to a target FSR level of $0.05$, while the right panel corresponds to a target FSR level of $0.1$.}

    \label{fig::simu3_results}
\end{figure}

In the second simulation, we generate \( N=2000 \) effects from a mixture of three normal distributions with equal proportions, and the covariances of observations are identity matrices. The three prior covariance matrices are given in the appendix (\ref{simu2:cov}). They are no longer rank-1, and the prior is (in a sense) mis-specified if we set the prior covariance to be rank-1. We do not assume the rank of prior covariance matrices and obtain estimated covariance matrices by running ED (or TEEM, see appendix (\ref{simu2:cov})) in unconstrained mode. We then vary the rank of the three estimated covariance matrices from 1 to 6 by truncating their eigenvalues at 0. The goal is to mimic the situation when the estimated covariance matrices are low-rank in the unconstrained mode. The rank-modified covariance matrices are given as input to the mash function. The plot in Figure \ref{fig::simu3_results} shows the FSR control at target levels \( \alpha=0.05 \) and \( \alpha=0.1 \). Our method improves the FSR control significantly and controls FSR uniformly at different ranks.

\section{Discussion}

In this paper, we study the relationship between lfsr and the rank of covariance matrices in the prior and provide detailed results for the rank-1 prior. The low-rank prior covariance matrices impose constraints on the effects and the lfsr. The rank-1 prior is the most extreme case, in which the lfsr of all conditions in one sample are exactly the same. Our proposed solution, using a full-rank covariance matrix, ensures that the lfsr of each condition can vary without additional constraints, and we have demonstrated the effectiveness of this solution in the simulations. We assume that the true effects are not exactly 0. This assumption ensures that the lfsr is at most 0.5—if we were to guess the sign randomly. If the true effects are allowed to be 0, then the lfsr can be at most 1. Since the estimated covariance matrices can never have exactly zero entries, the FSR is much harder to control.

In practice, the best approach to estimating prior covariance matrices is likely to initialize the algorithm with a full-rank covariance and then adjust the estimated ones to ensure they remain full-rank. On the other hand, the rank-1 prior covariance matrix is easier to interpret and visualize in real applications. One can simply plot the vector $\bs u$ and immediately observe the relative scale of all conditions. Rank-1 priors also make the computation much less intensive (especially for the matrix inverse) and can noticeably increase the speed of model fitting. In this sense, one may prefer to retain rank-1 matrices in the prior. We have shown in the second simulation that even in cases of prior misspecification—where the true prior covariances are not rank-1, but we only include a rank-1 prior in the model fitting—our method can still significantly reduce FSR inflation. The potential inflation of FSR due to prior mis-specification is not considered in this paper and is beyond its scope. The estimation uncertainty in the estimated prior is also not addressed in this paper; an approach to account for estimation error is to obtain uncertainty quantification of the estimated lfsr. See \citet{ignatiadis2021confidence, xie2022discussion} for more details. It would be also interesting to apply the method to applications such as differential expression \citep{squair2021confronting}, cell type deconvolution \citep{xie2022robust}, and single-cell studies \citep{yazar2022single}.

\bibliography{lfsr_rank}

\appendix
\section{Definitions of lfsr}
\label{appendix:lfsr_def}

The lfsr defined with respect to an estimator is
\begin{equation}\label{def:lfsr_estimator}
    lfsr = p(\mu\times\hat{\mu}(x)\leq 0|x).
\end{equation}

\begin{figure}
    \centering
    \includegraphics[scale=0.5]{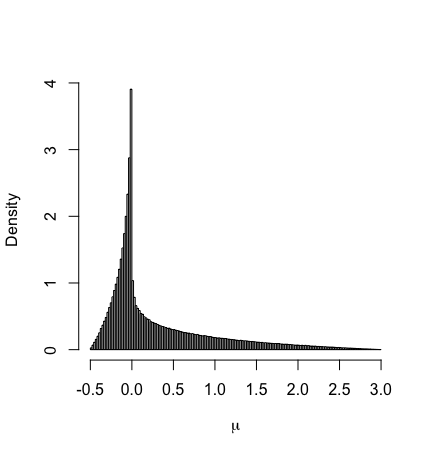}
    \caption{Histogram of a million draws of $\mu$}
    \label{fig:lfsr_def}
\end{figure}

Possible estimators of $\mu$ are $x$, $\E(\mu|x)$ or the posterior mode. They may give different estimates of the effect sign. For example, if $x\sim N(\mu,1)$ and $\mu\sim g(\cdot)$ where $g(\cdot)$ is a normal or mixture of normal distributions, $x$ and $\E(\mu|x)$ have the same sign. In general, whether their signs coincide depends on the likelihood and prior. Note that the definitions (\ref{def:lfsr_stephens}) and (\ref{def:lfsr_estimator}) do not always give the same results. If the effect sign was estimated positive based on posterior mean, it does not implicate $p(\mu\leq 0|x)$(given by definition (\ref{def:lfsr_estimator})) is the smaller one. The example below shows a case where the two definition differ. Assume $x\sim N(\mu,1)$ and $\mu$ follows a mixture of 200 half uniform distributions with equal probabilities. The half uniform distributions are either $[a_k,0]$ or $[0,b_k]$ where $k=1,2,...,100$, $a_k\in[-0.5,-0.01]$ and $b_k\in [0.01,3]$. The distribution of $\mu$ is thus unimodal at 0 and Figure \ref{fig:lfsr_def} shows a histogram of a million draws of $\mu$. The posterior results were obtained using ashr package. Given $x = -1$, the posterior mean of $\mu$ is $\E(\mu|x) = 0.0034$ so we claim the sign of $\mu$ to be positive. The lfsr based on definition (\ref{def:lfsr_estimator}) is $p(\mu\leq 0|x) = 0.70$ while it is $0.3$ based on (\ref{def:lfsr_stephens}). If we were to declare the sign of effect based on $x$ which is negative, then two definition would give the same lfsr. When $x=0.05$, the posterior mean of $\mu$ is $\E(\mu|x) = 0.14$, and $p(\mu\leq 0|x) = 0.57$. Again the two definition give different lfsr.

The lfsr applies even if the null distribution of effect is unspecified(e.g., we need not knowing which one is the null distribution in a mixture prior). For lfdr, due to its definition of $\mu$ being exactly 0(or the effect is null), we have to specify the null distribution in the model. For example the two-group model assumes a mixture prior of $\mu$,
\begin{equation}
    \mu\sim \pi_0 g_0(\mu) + \pi_1g_1(\mu),
\end{equation}
where $g_0(\mu)$ is the density of $\mu$ under the null hypothesis. The marginal distribution of $x$ is then $x\sim \pi_0 f_0(x) + \pi_1 f_1(x)$. The lfdr in the two-group model is $lfsr = \pi_0f_0(x)/f(x)$, which requires explicitly knowing $f_0$ is the null density. We can see from the definition of lfsr that it does not require specifying the null density. This difference seems trivial in the univariate model. But this property of lfsr makes it straightforward to extend the widely used univariate empirical Bayes multiple testing procedure to higher dimension. In multivariate settings, each sample has effects of multiple conditions and it's nontrivial to specify the null distribution especially when the prior distributions are estimated from data. 

\section{Derivation of Fisher information matrix of multivariate normal distribution}
\label{appendix:infomat}

Assume 
\begin{equation}
    \bs x_i\sim N(\bs 0, \bs U + \bs D + \bs V_i),
\end{equation}
where $\bs V_i\in\mathbb{R}^{R\times R}$ is a known matrix, $\bs D$ is a diagonal matrix and $\bs U$ is an unstructured covariance matrix. The goal is to derive the Fisher information matrix of parameters in $\bs U, \bs D$.

Denote $\bs T_i = \bs U + \bs D + \bs V_i$, and $\bs\theta = (\bs u,\bs \sigma^2)^T$, where $\bs u = vec(\bs U)$ and contains only unique elements so $\bs u\in\mathbb{R}^{R(R+1)/2}$, and $\bs\sigma^2 = (\sigma^2_1,...,\sigma^2_R)$. Note that elements of $\bs T_i$ are functions of $\bs\theta$ and for simplicity we write $\bs T_i$ instead of $\bs T_i(\bs\theta)$.

The log likelihood of $\bs x_i$ is 
\begin{equation}
    l(\bs\theta;\bs x_i) = -\frac{1}{2}\log|\bs T_i| - \frac{1}{2}\bs x_i^T\bs T_i^{-1}\bs x_i.
\end{equation}

The gradient of $l(\bs\theta;\bs x_i)$ with respect to the $j$th entry of $\bs\theta$ gives the score function 
\begin{equation}
    \frac{\partial l(\bs\theta;\bs x_i)}{\partial\theta_j} = -\frac{1}{2}\tr(\bs T_i^{-1}\frac{\partial\bs T_i}{\partial\theta_j})+ \frac{1}{2}\bs x_i^T\bs T_i^{-1}\frac{\partial\bs T_i}{\partial\theta_j}\bs T_i^{-1}\bs x_i.
\end{equation}

The second order derivative of $l(\bs\theta;\bs x_i)$ with respect to $\bs\theta$ gives the Hessian, 
\begin{equation}
\begin{split}
    \frac{\partial^2l(\bs\theta;\bs x_i)}{\partial\theta_j\partial\theta_k} = & -\frac{1}{2}\tr\left(\bs T_i^{-1}\frac{\partial^2\bs T_i}{\partial\theta_j\partial\theta_k} - \bs T_i^{-1}\frac{\partial\bs T_i}{\partial\theta_j}\bs T_i^{-1}\frac{\partial\bs T_i}{\partial\theta_k}\right)
    \\
    & +\frac{1}{2}\bs x_i^T\bs T_i^{-1}\left(\frac{\partial^2\bs T_i}{\partial\theta_j\partial\theta_k}  - 2\frac{\partial\bs T_i}{\partial\theta_j}\bs T_i^{-1}\frac{\partial\bs T_i}{\partial\theta_k}\right)\bs T_i^{-1}\bs x_i.
\end{split}
\end{equation}

The Fisher information matrix is given by the expectation of the negative Hessian,

\begin{equation}
 -\E(\frac{\partial^2l(\bs\theta;\bs x_i)}{\partial\theta_j\partial\theta_k}) = \frac{1}{2}\tr\left(\bs T_i^{-1}\frac{\partial\bs T_i}{\partial\theta_j}\bs T_i^{-1}\frac{\partial\bs T_i}{\partial\theta_k}\right).
\end{equation}

We denote the Fisher information matrix as $\bs I(\bs\theta)$, and it's $j,k$th element is 
\begin{equation}
    \bs I_{jk}(\bs\theta) = \frac{1}{2}\sum_i\tr\left(\bs T_i^{-1}\frac{\partial\bs T_i}{\partial\theta_j}\bs T_i^{-1}\frac{\partial\bs T_i}{\partial\theta_k}\right).
\end{equation}

The asymptotic variance of $\hat{\bs\theta}^{mle}$ is given by $I(\bs\theta)^{-1}$, and can be estimated by $I(\hat{\bs\theta})^{-1}$.

We can partition the information matrix into the following four blocks

\begin{equation}
\begin{split}
    \bs I(\bs\theta) &=  \left[ 
   \setlength\arraycolsep{20pt}
   \renewcommand{\arraystretch}{2}
\begin{array}{c | c} 
   \frac{\partial^2 l(\bs\theta)}{\partial\bs u \partial\bs u^T}& \frac{\partial^2 l(\bs\theta)}{\partial\bs u \partial(\bs\sigma^2)^T}  \\ 
  \hline 
  \frac{\partial^2 l(\bs\theta)}{\partial\bs{\sigma^2}^T \partial\bs u} & \frac{\partial^2 l(\bs\theta)}{\partial\bs{\sigma^2}^T \partial\bs{\sigma^2}}
\end{array} 
\right] 
\\
& = \left[ 
   \setlength\arraycolsep{20pt}
   \renewcommand{\arraystretch}{2}
\begin{array}{c | c} 
   \bs A & \bs B \\ 
  \hline 
  \bs B^T & \bs C
\end{array} 
\right] 
\end{split}
\end{equation}

We order the elements in $\bs u$ as the unique elements in each row of matrix $\bs U$, i.e. 
\begin{equation}
    \bs u = (u_{11},u_{12},...,u_{1R},u_{22},...,u_{2R},...,u_{RR}).
\end{equation}

The matrix $\bs A$ of dimension $\frac{R(R+1)}{2}\times \frac{R(R+1)}{2}$ is 
\begin{equation}
    \bs A = \begin{pmatrix}
    
    \frac{\partial^2l(\bs\theta)}{\partial^2 u_{11}} 
    & \frac{\partial^2l(\bs\theta)}{\partial u_{11}\partial u_{12}}
    &\cdots
    & \frac{\partial^2l(\bs\theta)}{\partial u_{11}\partial u_{RR}}
    \vspace{2mm}
    \\
    
    \frac{\partial^2l(\bs\theta)}{\partial u_{12}\partial u_{11}}
    & \frac{\partial^2l(\bs\theta)}{\partial^2 u_{12}}
    &\cdots
    &\frac{\partial^2l(\bs\theta)}{\partial u_{12}\partial u_{RR}}
    \vspace{2mm}
    \\
    \vdots&\vdots&\vdots&\vdots
    \vspace{2mm}
    \\
    \frac{\partial^2l(\bs\theta)}{\partial u_{RR}\partial u_{11}}
    & \frac{\partial^2l(\bs\theta)}{\partial u_{RR}\partial u_{12}}
    &\cdots
    &\frac{\partial^2l(\bs\theta)}{\partial^2 u_{RR}}
    \end{pmatrix}.
\end{equation}

The matrix $\bs B$ of dimension $\frac{R(R+1)}{2}\times R$ is 

\begin{equation}
    \bs B = 
    \begin{pmatrix}
    \frac{\partial^2l(\bs\theta)}{\partial u_{11}\partial\sigma^2_1} 
    & \frac{\partial^2l(\bs\theta)}{\partial u_{11}\partial\sigma^2_2} 
    &\cdots
    &\frac{\partial^2l(\bs\theta)}{\partial u_{11}\partial\sigma^2_R} 
    \vspace{2mm}
    \\
    \frac{\partial^2l(\bs\theta)}{\partial u_{12}\partial\sigma^2_1} 
    & \frac{\partial^2l(\bs\theta)}{\partial u_{12}\partial\sigma^2_2} 
    &\cdots
    &\frac{\partial^2l(\bs\theta)}{\partial u_{12}\partial\sigma^2_R} 
    \vspace{2mm}
    \\
    \vdots&\vdots&\vdots&\vdots
    \vspace{2mm}
    \\
    \frac{\partial^2l(\bs\theta)}{\partial u_{RR}\partial\sigma^2_1} 
    & \frac{\partial^2l(\bs\theta)}{\partial u_{RR}\partial\sigma^2_2} 
    &\cdots
    &\frac{\partial^2l(\bs\theta)}{\partial u_{RR}\partial\sigma^2_R} 
    \end{pmatrix}.
\end{equation}

The matrix $\bs C$ of dimension $R\times R$ is 
\begin{equation}
\bs C = 
    \begin{pmatrix}
    \frac{\partial^2l(\bs\theta)}{\partial^2\sigma^2_1}
    & \frac{\partial^2l(\bs\theta)}{\partial \sigma^2_1\partial\sigma^2_2}
    &\cdots
    &\frac{\partial^2l(\bs\theta)}{\partial \sigma^2_1\partial\sigma^2_R}
    \vspace{2mm}
    \\
    \frac{\partial^2l(\bs\theta)}{\partial\sigma^2_2\partial\sigma^2_1}
    & \frac{\partial^2l(\bs\theta)}{\partial^2\sigma^2_2}
    &\cdots
    &\frac{\partial^2l(\bs\theta)}{\partial \sigma^2_2\partial\sigma^2_R}
    \vspace{2mm}
    \\
    \vdots&\vdots&\vdots&\vdots
    \vspace{2mm}
    \\
    \frac{\partial^2l(\bs\theta)}{\partial\sigma^2_R\partial\sigma^2_1}
    & \frac{\partial^2l(\bs\theta)}{\partial \sigma^2_R\partial\sigma^2_2}
    &\cdots
    &\frac{\partial^2l(\bs\theta)}{\partial^2\sigma^2_R}
    \end{pmatrix}.
\end{equation}

Let $\bs e_j$ denote a vector that the $j$th entry is 1 and others are 0. The following derivatives hold 
\begin{equation}
    \begin{split}
        & \frac{\partial\bs T_i}{\partial u_{rr}} = \frac{\partial\bs U}{\partial u_{rr}} = \bs e_r\bs e_r^T,
        \\
        & \frac{\partial\bs T_i}{\partial u_{r_1r_2}} = \frac{\partial\bs U}{\partial u_{r_1r_2}} = \bs e_{r_1}\bs e_{r_2}^T + \bs e_{r_2}\bs e_{r_1}^T,
        \\
        & \frac{\partial\bs T_i}{\partial \sigma^2_r} = \frac{\partial\bs D}{\partial \sigma^2_r} = \bs e_r\bs e_r^T.
    \end{split}
\end{equation}

\subsection{Calculation of matrix $\bs A$:}

Denote $\tilde{\bs S}_i = \bs T_i^{-1}$, with column vectors $\tilde{\bs  s}_{i1},...,\tilde{\bs  s}_{iR}$.

Denote the $(r_1,r_2)$th entry of matrix $\tilde{\bs S}^{-1}$ as $\tilde{s}_{i,r_1r_2}$.

We denote the entries of matrix $\bs A$ as $a_{rj_r,lk_l}$, for $r,l\in [R]$,  $j_r = r,r+1,...,R-r+1$, and $k_l = l,l+1,...,R-l+1$. Each entry is given by 
\begin{equation}
    a_{rj_r,lk_l} = \frac{1}{2}\sum_i\tr\left(\bs T_i^{-1}\frac{\partial\bs T_i^{-1}}{\partial u_{rj_r}}\bs T_i^{-1}\frac{\partial\bs T_i^{-1}}{\partial u_{lk_l}}\right).
\end{equation}

The matrix $\bs A$ is symmetric since $a_{rj_r,lk_l} = a_{lk_l,rj_r}$. We consider the following cases: 

\begin{itemize}
    \item When $r=j_r$, $l=k_l$, 
    \begin{equation}
        \begin{split}
            a_{rr,ll} &= \frac{1}{2}\sum_i\tr\left(\bs T_i^{-1}\frac{\partial\bs U}{\partial u_{rr}}\bs T_i^{-1}\frac{\partial\bs U}{\partial u_{ll}}\right)
            \\
            &= \frac{1}{2}\sum_i \tr\left(\bs T_i^{-1}\bs e_r\bs e_r^T\bs T_i^{-1}\bs e_l\bs e_l^T\right)
            \\
            &= \frac{1}{2}\sum_i \bs e_l^T\left(\bs T_i^{-1}\bs e_r\bs e_r^T\bs T_i^{-1}\right)\bs e_l
            \\
            & = \frac{1}{2}\sum_i (\tilde{s}_{i,rl})^2. 
        \end{split}
    \end{equation}
    \item When $r=j_r, l\neq k_l$, 
    \begin{equation}
        \begin{split}
            a_{rr,lk_l} &= \frac{1}{2}\sum_i\tr\left(\bs T_i^{-1}\frac{\partial\bs U}{\partial u_{rr}}\bs T_i^{-1}\frac{\partial\bs U}{\partial u_{lk_l}}\right)
            \\
            &= \frac{1}{2}\sum_i \tr\left(\bs T_i^{-1}\bs e_r\bs e_r^T\bs T_i^{-1}(\bs e_l\bs e_{k_l}^T+\bs e_{k_l}\bs e_l^T)\right)
            \\
            &= \frac{1}{2}\sum_i \left(\bs e_{k_l}^T\left(\bs T_i^{-1}\bs e_r\bs e_r^T\bs T_i^{-1}\right)\bs e_l + \bs e_{l}^T\left(\bs T_i^{-1}\bs e_r\bs e_r^T\bs T_i^{-1}\right)\bs e_{k_l}\right)
            \\
            & = \sum_i \tilde{s}_{i,rl}\tilde{s}_{i,rk_l}. 
        \end{split}
    \end{equation}
    \item When $r\neq j_r, l\neq k_l$, 
    \begin{equation}
        \begin{split}
            a_{rj_r,lk_l} &= \frac{1}{2}\sum_i\tr\left(\bs T_i^{-1}\frac{\partial\bs U}{\partial u_{rj_r}}\bs T_i^{-1}\frac{\partial\bs U}{\partial u_{lk_l}}\right)
            \\
            &= \frac{1}{2}\sum_i \tr\left(\bs T_i^{-1}(\bs e_r\bs e_{j_r}^T+\bs e_{j_r}\bs e_r^T)\bs T_i^{-1}(\bs e_l\bs e_{k_l}^T+\bs e_{k_l}\bs e_l^T)\right)
            \\
            & = \sum_i (\tilde{s}_{i,lj_r}\tilde{s}_{i,rk_l} + \tilde{s}_{i,rl}\tilde{s}_{i,j_r k_l}). 
        \end{split}
    \end{equation}
\end{itemize}

\subsection{Calculation of matrix $\bs B$}

Denote the entries of matrix $\bs B$ as $b_{rj_r,l}$, for $r,l\in[R]$, $j_r = r,r+1,...,R-r+1$. Each entry is given by 

\begin{equation}
    b_{rj_r,l} = \frac{1}{2}\sum_i\tr\left(\bs T_i^{-1}\frac{\partial\bs U}{\partial u_{rj_r}}\bs T_i^{-1}\frac{\partial\bs D}{\partial \sigma^2_l}\right).
\end{equation}

We consider the following cases:

\begin{itemize}
    \item When $r=j_r$, 
    \begin{equation}
        \begin{split}
            b_{rr,l} &= \frac{1}{2}\sum_i\tr\left(\bs T_i^{-1}\frac{\partial\bs U}{\partial u_{rr}}\bs T_i^{-1}\frac{\partial\bs D}{\partial \sigma^2_l}\right)
           \\
           &= \frac{1}{2}\sum_i\tr\left(\bs T_i^{-1}(\bs e_r\bs e_r^T )\bs T_i^{-1}(\bs e_l\bs e_l^T )\right)
           \\
           &= \frac{1}{2}\sum_i (\tilde{s}_{i,rl})^2.
        \end{split}
    \end{equation}
    \item When $r\neq j_r$, 
    \begin{equation}
        \begin{split}
            b_{rj_r,l} &= \frac{1}{2}\sum_i\tr\left(\bs T_i^{-1}\frac{\partial\bs U}{\partial u_{rj_r}}\bs T_i^{-1}\frac{\partial\bs D}{\partial \sigma^2_l}\right)
           \\
           &= \frac{1}{2}\sum_i\tr\left(\bs T_i^{-1}(\bs e_r\bs e_{j_r}^T+\bs e_{j_r}\bs e_{r}^T )\bs T_i^{-1}(\bs e_l\bs e_l^T )\right)
           \\
           &= \sum_i (\tilde{s}_{i,rl}\tilde{s}_{i,j_rl}).
        \end{split}
    \end{equation}
\end{itemize}





In the special case where $\bs D=  \sigma^2\bs I$, $\bs B = \bs b$ is a vector of length $R(R+1)/2$, and 

\begin{itemize}
    \item When $r=j_r$, 
    \begin{equation}
        \begin{split}
            b_{rr} &= \frac{1}{2}\sum_i\tr\left(\bs T_i^{-1}\frac{\partial\bs U}{\partial u_{rr}}\bs T_i^{-1}\right)
           \\
           &= \frac{1}{2}\sum_i\tr\left(\bs T_i^{-1}(\bs e_r\bs e_r^T )\bs T_i^{-1}\right)
           \\
           &= \frac{1}{2}\sum_i\sum_{l=1}^R (\tilde{s}_{i,rl})^2.
        \end{split}
    \end{equation}
    \item When $r\neq j_r$, 
    \begin{equation}
        \begin{split}
            b_{rj_r} &= \frac{1}{2}\sum_i\tr\left(\bs T_i^{-1}\frac{\partial\bs U}{\partial u_{rj_r}}\bs T_i^{-1}\right)
           \\
           &= \frac{1}{2}\sum_i\tr\left(\bs T_i^{-1}(\bs e_r\bs e_{j_r}^T+\bs e_{j_r}\bs e_{r}^T )\bs T_i^{-1}\right)
           \\
           &= \sum_i\sum_{l=1}^R (\tilde{s}_{i,rl}\tilde{s}_{i,j_rl}).
        \end{split}
    \end{equation}
\end{itemize}

\subsection{Calculation of matrix $\bs C$}

Denote the entries of $\bs C$ as $c_{rl}$, for $r,l\in[R]$. Each entry of $\bs C$ is given by 
\begin{equation}
    \begin{split}
        c_{rl} &= \frac{1}{2}\sum_i\tr\left(\bs T_i^{-1}\frac{\partial\bs D}{\partial \sigma^2_r}\bs T_i^{-1}\frac{\partial\bs D}{\partial \sigma^2_l}\right)
        \\
        &= \frac{1}{2}\sum_i\tr\left(\bs T_i^{-1}(\bs e_r\bs e_r^T)\bs T_i^{-1}(\bs e_l\bs e_l^T)\right)
        \\
        & = \frac{1}{2}\sum_i \tilde{s}_{i,rl}^2
    \end{split}
\end{equation}

In the special case where $\bs D = \sigma^2\bs I$, $\bs C = c$ is a scalar,  and 

\begin{equation}
    \begin{split}
        c &= \frac{1}{2}\sum_i \tr\left(\bs T_i^{-1}\bs T_i^{-1}\right)
        \\
        &= \frac{1}{2}\sum_i \sum_r\sum_l \tilde{s}_{i,rl}^2.
    \end{split}
\end{equation}

\subsection{Inverse Fisher Information Matrix}

The variance of $\bs{\sigma}^2$ is given by 
\begin{equation}
    (\bs C - \bs B^T\bs A^{-1}\bs B)^{-1}.
\end{equation}

Let's consider the case $\bs D = \sigma^2\bs I$ so $\var(\hat{\sigma}^2) = \frac{1}{c-\bs b^T\bs A^{-1}\bs b}$. The difficulty of obtaining $\var(\hat{\sigma}^2)$ is from formulating $\bs b$, $\bs A$ and get the inverse of $\bs A$.

Since the Fisher information matrix is always positive semidefinite and we have assumed it's nonsingular, we have $\bs b^T\bs A^{-1}\bs b> 0$, and 
\begin{equation}
    \var(\hat{\sigma}^2) = \frac{1}{c-\bs b^T\bs A^{-1}\bs b} > \frac{1}{c}.
\end{equation}

We can use $1/c$ as a lower bound of $\var(
\hat{\sigma}^2)$ in practice(if not evaluating $\bs A$).

For a general $\bs D$, we are interested in the diagonal of  $\cov(\hat{\bs\sigma}^2) = (\bs C - \bs B^T\bs A^{-1}\bs B)^{-1}$. 

\begin{lemma}\label{lemma:invdiag}
Let $\bs X$ be a positive definite matrix, then $(\bs X^{-1})_{jj}\geq \frac{1}{\bs X_{jj}}$.
\end{lemma}
\begin{proof}
The eigen-decompostion of $\bs X$ is $\bs X = \bs Q\bs\Lambda\bs Q^T$. Then 
\[\bs X_{jj} = \bs q_j^T\bs\Lambda\bs q_j = \sum_kq_{jk}^2\lambda_k,\]
and 
\[(\bs X)^{-1}_{jj} = \bs q_j^T\bs\Lambda^{-1}\bs q_j = \sum_kq_{jk}^2/\lambda_k.\]

By Cauchy–Schwarz inequality,
\begin{equation*}
    \bs X_{jj}(\bs X)^{-1}_{jj} \geq (\sum_k q_{jk})^2 = 1. 
\end{equation*}
\end{proof}

Since by assumption $\bs C - \bs B^T\bs A^{-1}\bs B$ is positive definite, by lemma (\ref{lemma:invdiag}), we have 

\begin{equation}
    (\bs C - \bs B^T\bs A^{-1}\bs B)^{-1}_{rr}\geq 1/(\bs C - \bs B^T\bs A^{-1}\bs B)_{rr} = \frac{1}{\bs C_{rr} - \bs b_r^T\bs A^{-1}\bs b_r}\geq \frac{1}{\bs C_{rr}}.
\end{equation}

We summarize the results below.

\begin{enumerate}\label{MVN:lowerbound}
    \item When $\bs V_i = \bs I$, 
    \begin{itemize}
        \item if $\bs D = \sigma^2 \bs I$, $\var(\hat{\sigma}^2)\geq\frac{2}{N\tr((\bs U + \bs I)^{-1}(\bs U + \bs I)^{-1})}$.
        \item if $\bs D = \diag(\bs{\sigma}^2)$, $\var(\hat{\sigma}^2_r)\geq \frac{2}{N((\bs U + \bs I)^{-1}_{rr})^2}$.
    \end{itemize}
    \item When $\bs V_i = \bs V$, 
    \begin{itemize}
        \item if $\bs D = \sigma^2 \bs I$, $\var(\hat{\sigma}^2)\geq\frac{2}{N\tr((\bs U + \bs V)^{-1}(\bs U + \bs V)^{-1})}$.
        \item if $\bs D = \diag(\bs{\sigma}^2)$, $\var(\hat{\sigma}^2_r)\geq \frac{2}{N((\bs U + \bs V)^{-1}_{rr})^2}$.
    \end{itemize}
    \item When $\bs V_i$ varies with samples, 
    \begin{itemize}
        \item if $\bs D = \sigma^2 \bs I$, $\var(\hat{\sigma}^2)\geq\frac{2}{\sum_i \tr((\bs U + \bs V_i)^{-1}(\bs U + \bs V_i)^{-1})}$.
        \item if $\bs D = \diag(\bs{\sigma}^2)$, $\var(\hat{\sigma}^2_r)\geq \frac{2}{\sum_i((\bs U + \bs V_i)^{-1}_{rr})^2}$.
    \end{itemize}
\end{enumerate}

If $\bs V_i$ is diagonal and $\bs U = 0$, then $\var(\hat{\sigma}^2_r)\geq 2(\sum_i((\bs V_{i}^{-1})_{rr})^{2})^{-1}$.


\section{Lower bound of variance}
\label{appendix:lb}
We start with a simple one dimensional case. Assume $x_i\sim N(0,1+\sigma^2)$, our goal is to estimate $\sigma^2$ and possibly set $\hat{\sigma}^2$ to be the largest value of it consistent with data. 

The log likelihood of $l(\sigma^2)$ is 
\begin{equation}
    l(\sigma^2) = -\frac{n}{2}\log(1+\sigma^2) - \frac{ns^2}{2(1+\sigma^2)},
\end{equation}
where $s^2 = \sum_ix_i^2/n$, the MLE of $(1+\sigma^2)$.

Then we have 
\begin{equation}
    l(\sigma^2) - l(0) = -\frac{n}{2}\left(\log(1+\sigma^2) - \frac{\sigma^2}{1+\sigma^2}s^2\right).
\end{equation}

When $s^2=1$(truncated) or equivalently $\hat{\sigma}^2 = 0$, and $\sigma^2$ is very small, using the approximation $\log(1+\sigma^2)\approx\sigma^2$, we have 

\begin{equation}
    l(\sigma^2) - l(0) \approx -\frac{n}{2}\frac{(\sigma^2)^2}{1+\sigma^2} \approx -\frac{n}{2}(\sigma^2)^2.
\end{equation}

Ae rule of thumb is that 2 units drop in log likelihood gives $95\%$ confidence intervals. For example, $l(\hat{\theta})-2\approx l(\hat{\theta}+2*se(\hat{\theta}))$.

Solving for $l(\sigma^2) - l(0) = -2$, we have $\sigma^2 \approx \frac{2}{\sqrt{n}}$. This suggests that we can set the lower bound of $\sigma^2$ to be $\frac{2}{\sqrt{n}}$. Note that this lower bound is based on Wilks' theorem.

More generally, assume
\begin{equation}
    x_i\sim N(0,\sigma^2 + v_i^2),
\end{equation}
where $v_i^2$ is known and $\sigma^2$ is to be estimated. The MLE of $\sigma^2$ is the solution to the following fixed point problem 
\begin{equation}
    \sigma^2 = \frac{\sum_i w_i^2(x_i^2-v_i^2)}{\sum_iw_i^2},
\end{equation}
where $w_i = \frac{1}{v_i^2+\sigma^2}$.

The variance of $\hat{\sigma}^2$ is given by the inverse fisher information, 
\begin{equation}
    var(\hat{\sigma}^2) = 2(\sum_iw_i^2)^{-1}.
\end{equation}

When $\sigma^2$ is small and $\hat{\sigma}^2 = 0$, we can set $\hat{\sigma}^2$ to be $2\sqrt{2}\sqrt{\frac{1}{\sum_i (v_i^2)^{-2}}}$, which is the upper bound of the $95\%$ Wald confidence interval. When $v_i^2 = 1$ for all samples, the upper bound is $\sqrt{2}$ times the lower bound based on the rule of 2.

The lower bound of variance of $\widehat{\bs D}$ in multivariate normal distribution is discussed in above section. Similarly we can get the lower bound of the variance of $\text{diag}(\widehat{\bs U})$ and is summarized below:

\begin{enumerate}\label{MVN:lowerboundU}
    \item When $\bs V_i = \bs I$, $\var(\hat{u}^2_r)\geq \frac{2}{N}$;
    
    \item When $\bs V_i = \bs V$, $\var(\hat{u}^2_r)\geq \frac{2}{N( \bs V^{-1}_{rr})^2}$;
\item When $\bs V_i$ varies with samples, $\var(\hat{u}^2_r)\geq \frac{2}{\sum_i (\bs V_i^{-1})_{rr}^2}$.

\end{enumerate}

In mixture case, we consider $\bs x_i\sim\sum_k\pi_k N(\bs 0,\bs U_k+\bs I)$ and define $\bs T_k = \bs U_k+\bs I$. When estimating $\bs U_k$ using EM algorithm, the last iteration gives 
\begin{equation}
    \hat{\bs T}_k^{\text{mle}} = \frac{\sum_i\gamma_{ik}\bs x_i\bs x_i^T}{n_k},
\end{equation}
 where $\text{mle}$ denotes the maximum likelihood estimator, $\gamma_{ik}$ are posterior probabilities of $\bs x_i$ coming from mixture component $k$ and $n_{k} = \sum_i\gamma_{ik}$. (In practice one would use a modified $\hat{\bs T}_k^{\text{mle}}$ by truncating eigenvalues to ensure $\bs U_k$ is a well-defined covariance matrix, and $\hat{\bs U}_k^{\text{mle}} = \hat{\bs T}_k^{\text{mle}} - \bs I$.) Most of the posterior probabilities $\gamma_{ik}$ should be close to either 0 or 1 at the last iteration of EM. The reason is that assuming the model is correct, the posteriors should concentrate on the true mixture component assignments. The number $n_k$ can be regarded as the effective sample size of mixture component $k$, i.e. the number of $\bs x_i$ that are from the mixture $k$. Therefore, we have the following approximation
 \begin{equation}
     \hat{\bs T}_k^{\text{mle}} = \frac{\sum_i\gamma_{ik}\bs x_i\bs x_i^T}{n_k}\approx \frac{1}{n_k}\sum_{i\in\mathcal{K}}\bs x_i\bs x_i^T,
 \end{equation}
 where $\mathcal{K}$ is an index set of samples from mixture component $k$, so for $i\in\mathcal{K}$, $\bs x_i\sim N(\bs 0,\bs U_k)$. We can regard $n_k$ as an effective sample size of the mixture component $k$ and  obtain the lower bound of the variances of $\bs U_k$(or $\bs D_k$) based on $n_k$. This applies when $\bs V_i = \bs I$. For general $\bs V_i$, we can use $\gamma_{ik}$ as weights in the  summation over samples when calculating lower bounds. For example, when $\bs V_i$ varies with samples, $\var(\hat{u}^2_{k,r})\geq \frac{2}{\sum_i\gamma_{ik}(\bs (V_i^{-1})_{rr})^2}$.
 
\section{Simulation details}
\label{simu2:cov}

\subsection{Three covariance matrices}

The first covariance matrix is 

\begin{equation}
    \begin{pmatrix} 
1 & 0.5 & 0 & 0 & 0 & 0\\
0.5& 1 & 0 & 0 & 0 & 0\\
 0 & 0 & 0.01 & 0 & 0 & 0\\
 0 & 0 & 0 & 0.01 & 0 & 0\\
 0 & 0 & 0 & 0 & 0.01 & 0\\
0 & 0 & 0 & 0 & 0 & 0.01
 \end{pmatrix}.
\end{equation}

 In the second one we swap the first $2\times 2$ matrix $\begin{pmatrix}1&0.5\\0.5&1\end{pmatrix}$ with the $\begin{pmatrix}0.01&0\\0&0.01\end{pmatrix}$ in the center then set the two 0.5 to be negative.
 
 In the third one we swap the first $2\times 2$ matrix $\begin{pmatrix}1&0.5\\0.5&1\end{pmatrix}$ with the $\begin{pmatrix}0.01&0\\0&0.01\end{pmatrix}$ in the bottom right.
 
 The three covariance matrices are multiplied by 3 for data generation.
 
 \subsection{Results using TEEM}
 
 \begin{figure}[h]
    \centering
    \includegraphics[scale=0.35]{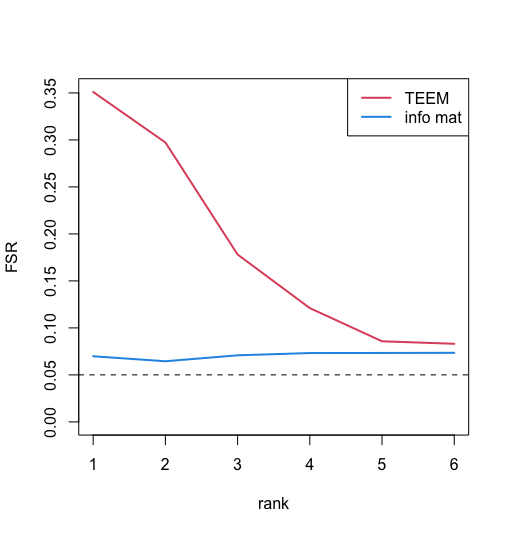}
    \caption{How FSR changes with the rank of estimated covariance matrices from TEEM. Target FSR level $0.05$.}
    \label{fig:simu2_fsr_teem}
\end{figure}
 
\medskip

\end{document}